\documentclass{ifacconf}
\usepackage{tikz}
\usepackage{graphicx,color} 
\usepackage{natbib} 
\usepackage{amsmath}
\usepackage{amssymb}
\usepackage{mathtools}

\newcommand{\RR}{\mathbb{R}}

\def\K{\mathcal{K}}

\def\e{\varepsilon}

\newtheorem{assumption}{Assumption}
\newtheorem{definition}{Definition}

\newtheorem{remark}{Remark}
\newtheorem{theorem}{Theorem}
\usepackage{float}
 
\usepackage{enumitem}
\setlist[itemize]{leftmargin=*}
\setlist[enumerate]{leftmargin=*}
\usepackage{comment}

\usepackage{array}

\newcommand{\fav}{f_{\text{av}}}
\begin{document}
\begin{frontmatter}

\title{Stability of Slow-Fast Nonlinear Dynamics: Non-Periodic Case}

\author[First]{G. Q. Bao Tran}
\author[First]{Daniel Liberzon}
\author[Second]{Hyungbo Shim}

\address[First]{Coordinated Science Laboratory, University of Illinois Urbana-Champaign, Urbana, IL 61801, USA\\(e-mail: \{baotran,liberzon\}@illinois.edu).}
\address[Second]{ASRI, Department of Electrical and Computer Engineering, Seoul National University, Seoul, South Korea (e-mail: hshim@snu.ac.kr).}

\begin{abstract}
We present sufficient conditions for the semi-global exponential stability of nonlinear systems whose dynamics have both slow and fast time variations. Unlike most existing results, the fast variation is non-periodic, thereby allowing a wider class of systems, especially switched systems with fast (non-periodic) switching and those with quasi-periodic variations; we therefore rely on general averaging to construct an average system. It is assumed that the average system admits a time-invariant equilibrium that is globally exponentially stable when the slow variation is frozen, i.e., remaining at a fixed value. This slow variation is allowed to be discontinuous in time, provided its total variation (flows and jumps) is bounded. The main result is illustrated using a nonlinear switched system with slow-fast non-periodic switching.
\end{abstract}

\end{frontmatter}

\section{Introduction}\label{sec:intro}
Consider a nonlinear system
\begin{equation}\label{eq:sys}
\dot{x}(t) = f(x(t),u_s(t),u_f(t/\e)),
\end{equation}
where $x \in \RR^n$ is the state, $\e > 0$ is a small parameter, and where $u_s \in \RR^m$ and $u_f \in \RR^l$ are slowly- and fast-varying signals modeling time variations in the dynamics. We would like to study sufficient conditions for the stability of system~\eqref{eq:sys}. When the dynamics are linear and only the slow variation is present, textbook references such as~\cite[Section 3.4]{ioannou1996robust} and~\cite[Section 9.6]{khalil} give exponential stability when the slow variation is sufficiently slow. When there are both slow and fast variations, informally, the main steps in investigating this class of systems involve averaging out $u_f$ first to obtain an \emph{average} system~\citep[Chapter 10]{khalil}, whose stability can be shown when $u_s$ is slow enough and $u_f$ is sufficiently fast, and then recover stability of the original system using perturbation theory.

In the literature, several results have been established regarding stability in such systems with slow and fast dynamics, as summarized in Table~\ref{tab_literature}. There are typically three main factors (or assumptions) to consider: 
\begin{itemize}[leftmargin=*,nosep]
\item whether $u_s,u_f$ are allowed to be discontinuous in time. Usually, $u_f$ can exhibit isolated jumps because it will be averaged out in the analysis. However, allowing $u_s$ to jump connects these results with those in switched systems like~\cite{hespanha1999,liberzonBook,shorten2007stability,liu2021integral,chitour2025dynamics} and the references therein. The discontinuity of $u_s$ is not considered in early works in this line of research~\citep{Peuteman2002,ChoiShimSeo}; 
\item whether the system has the same equilibrium with respect to the time variations of $u_s$ and $u_f$. In a simpler setting, if the equilibrium remains the same despite $u_s$ and $u_f$, then one can show exponential stability. This is referred to as the \emph{common} equilibrium case. On the other hand, if the equilibrium moves (with either $u_s$ or $u_f$), we can only achieve practical stability. This is also related to the rich literature on the stability of switched systems with multiple equilibria~\citep{mastellone,alpcan2010stability,makarenkov2018dwell,veer,hafez,yinhao} and the references therein;
\item whether the fast variation $u_f$ is periodic or not. If it is, the analysis is simplified thanks to a straightforward averaging method. However, if $u_f$ is non-periodic (as considered in this work), we must use a form of general averaging that is more intricate to analyze.
\end{itemize}

Several works on multi-time-scale systems can be found in the literature. For instance,~\cite{teel2003unified} analyzes slow and fast variations independently before interconnecting them, while~\cite{iannelli2004,teel2010averaging,sanfelice2011,abdelgalil2023multi} (and the references therein) study such behaviors within the frameworks of hybrid systems and singular perturbation theory. For the linear case, global exponential stability has been studied in a series of works~\citep{LibShimCDC22,ShimLibACM24,ShimLibNAHS25}. For nonlinear systems, semi-global exponential stability results have been obtained for systems of the form~\eqref{eq:sys}, where $u_s$ can jump, the origin is a constant equilibrium for both $u_s$ and $u_f$ without loss of generality, i.e., $f(0,\cdot,\cdot) = 0$, and $u_f$ is periodic~\citep{LibShimCDC24,LibShimTAC25}.~\cite{ChoiShimSeo} shows practical stability with non-periodic $u_f$ and moving equilibrium without treating discontinuities in $u_s$.

\begin{table*}[t]
\centering
\caption{Summary of results on slow-fast systems in chronological order.\label{tab_literature}}

\resizebox{\textwidth}{!}{
\renewcommand{\arraystretch}{1.35}
\begin{tabular}{|c|c|c|c|c|c|}
\hline
\textbf{Work} & \textbf{System dynamics} & \textbf{$u_s$ jumps?} & \textbf{Equilibrium} & \textbf{$u_f$ periodic?} & \textbf{Stability type} \\
\hline\hline
\cite{Peuteman2002}    & \begin{tabular}{c}written differently but equivalent to\\$\dot{x}(t) = f(x(t),u_s(t),u_f(t/\e))$\end{tabular} & no & common &  no & local, exponential\\\hline
\cite{ChoiShimSeo}    & $\dot{x}(t) = f(x(t),u_s(t),u_f(t/\e))$ & no & moving &  no & semi-global, practical\\\hline
\cite{LibShimCDC22}   & $\dot{x}(t) = (A(t) + B(t/\e))x(t)$  & yes & common  & yes & global, exponential\\\hline
\cite{ShimLibACM24}   & $\dot{x}(t) = (A(t) + B_s(t)B_f(t/\e))x(t)$  & yes & common  & yes & global, exponential\\\hline
\cite{Ahn}  & $\dot{x}(t) = f(x(t),u_s(t)) + g(x(t),u_f(t/\e))$ & yes & common  & yes & \begin{tabular}{c}semi-global, practical\\(but provably exponential)\end{tabular}\\\hline
\cite{ShimLibNAHS25}  & $\dot{x}(t) = (A(t) + B(t,t/\e))x(t)$ & yes & common  & yes & global, exponential\\\hline
\cite{LibShimCDC24,LibShimTAC25}   & 
\begin{tabular}{c}
$\dot{x}(t) = f(x(t),u_s(t),u_f(t/\e))$ \\ 
$\dot{x}(t) = (A(t) + B_s(t)B_f(t/\e))x(t)$
\end{tabular}
& yes & common  & yes & \begin{tabular}{c}semi-global, exponential\\global, exponential\end{tabular}\\\hline
This work & $\dot{x}(t) = f(x(t),u_s(t),u_f(t/\e))$ & yes & common & no & semi-global, exponential\\
\hline
\end{tabular}
}
\end{table*}

This work treats the common equilibrium case where $u_s$ can be discontinuous in time. Compared to our most recent work~\cite{LibShimTAC25}, it concerns a broader class of systems in the sense that $u_f$ is considered non-periodic, necessitating general averaging, which is harder to analyze~\citep[Section 10.6]{khalil}. Notably, this framework allows us to effectively deal with: i) switched systems with fast non-periodic switching, which are typically not common in the literature (see, e.g.,~\cite{mosta} which treats the linear case, or~\cite{pwm2025} for some PWM-based control methods), and ii) those with fast quasi-periodic variations that are rationally independent, which are not covered by~\cite{LibShimTAC25}. We illustrate the theoretical result using a nonlinear switched system with both slow and fast non-periodic switching.

\emph{Notations:} Let $\RR_{\geq 0} := [0,+\infty)$. Let $|\cdot|$ be the Euclidean vector norm and $\|\cdot\|$ be the induced matrix norm. Note that $\int\|\cdot\|$ is defined in Definition~\ref{def_tv}. The notion of class-$\K$ function is from~\citep[Definition 4.2]{khalil}.

\section{Main Result}
Consider system~\eqref{eq:sys}. Without loss of generality, we assume that this system has an equilibrium at $0$. 
\begin{assumption}[System~\eqref{eq:sys}]\label{ass_f}
The function $f$ is such that $f(0,\cdot,\cdot) = 0$ and is $C^1$. Moreover, its partial derivatives with respect to $x$ and $u_s$ are $C^1$ in $x$. 
\end{assumption}
To enable averaging, the following assumptions are made.
\begin{assumption}[Fast input~\citep{ChoiShimSeo}]\label{ass_fast_1}
The function $u_f$ is bounded, has finitely many discontinuities on any bounded interval, and is such that $f(x,u_s,u_f(\cdot))$ has a $C^1$ general average, i.e., for each $(x,u_s) \in \RR^n \times \RR^m$, there exists a limit~\citep[Section 10.6]{khalil} that is $C^1$
\begin{multline}\label{eq:ave}
\fav(x,u_s) := \lim_{T \to +\infty} \frac{1}{T}\int_t^{t+T} f(x,u_s,u_f(s))ds, \\ \forall t \geq 0.
\end{multline}
More formally, define the $C^1$ function
\begin{equation}
h(x,u_s,u_f) := f(x,u_s,u_f) -\fav(x,u_s).
\end{equation}
Assume that $h$ has zero average with convergence function $\sigma_D$, i.e., for each compact set $D \subset \RR^n \times \RR^m$, there is a strictly decreasing continuous function $\sigma_D: [0,+\infty) \to [0,+\infty)$ such that $\lim_{T \to +\infty}\sigma_D(T) = 0$ and
\begin{multline}\label{eq:sD}
\frac{1}{T}\left|\int_t^{t+T}h(x,u_s,u_f(s))ds\right|\leq \sigma_D(T), \\ \forall (x,u_s) \in D, \forall t,T\geq 0.
\end{multline}
\end{assumption}
Note that from Assumption~\ref{ass_f}, we have $\fav(0,\cdot)=0$, and thus $h(0,\cdot,\cdot)=0$.


\begin{remark}
When the fast input $u_f=(u_{f,1},u_{f,2},\ldots,u_{f,l})$ is composed of $l$ periodic components with \emph{rationally independent} periods $T_i$, $i=1,2,\ldots,l$, the combined fast signal is non-periodic but still quasi-periodic. It can be shown that in this special case, for each fixed $(x,u_s)$, the long-time average~\eqref{eq:ave} satisfies
\begin{multline}
\fav(x,u_s)
=
\frac{1}{T_1T_2\ldots T_l}
\int_0^{T_1}\int_0^{T_2}\ldots\int_0^{T_l}
\\f(x,u_s,(u_{f,1}(s_1),u_{f,2}(s_2)
\ldots,u_{f,l}(s_l)))ds_l\ldots ds_2 ds_1 .
\end{multline}
Therefore, the general average~\eqref{eq:ave} is time-invariant and coincides with the \emph{sequential} average over individual periods. This allows $\fav$ to be computed more easily.
\end{remark}

The following assumptions are made related to averaging. They are standard as made in~\cite[Section 10.6]{khalil}.
\begin{assumption}\label{ass_fast_2} 
The Jacobians $\frac{\partial h}{\partial x}$ and $\frac{\partial h}{\partial u_s}$ also have zero average with the same convergence function $\sigma_D$.\footnote{These maps are well defined because $\fav$ is $C^1$ and could in fact have different convergence functions. In the end, we take the largest one for all $h$, $\frac{\partial h}{\partial x}$, and $\frac{\partial h}{\partial u_s}$. For brevity, we denote them all as $\sigma_D$.}
\end{assumption}
We moreover assume that the average system
\begin{equation}
\dot{x} = \fav(x,u_s),
\end{equation}
which has $0$ as its equilibrium for any $u_s$, is globally exponentially stable for each \emph{frozen} $u_s$, i.e., stability when $u_s$ stays at a fixed admissible value. 
\begin{assumption}[Average system]\label{ass_average}
The function $u_s(\cdot)$ takes values in a compact and convex set $\Gamma \subset \RR^m$. There exists a $C^1$ function $V: \RR^n \times \RR^m \to \RR$ such that, for all $x \in \RR^n$ and for all $u_s \in \Gamma$, we have
\begin{subequations}\label{eq:V_av}
\begin{align}
c_1|x|^2 \leq V(x,u_s) & \leq c_2|x|^2,\label{eq:V_av_12}\\ \frac{\partial V}{\partial x}(x,u_s)\fav(x,u_s) &\leq -c_3 |x|^2,\label{eq:V_av_3}\\
\left|\frac{\partial V}{\partial u_s}(x,u_s)\right|&\leq c_4 |x|^2,\label{eq:V_av_4}\\
\left|\frac{\partial V}{\partial x}(x,u_s)\right|&\leq c_5 |x|,\label{eq:V_av_5}
\end{align}
\end{subequations}
for some $c_i >0$, $i=1,2,\ldots,5$.
\end{assumption}
Note that~\cite[Lemma 9.8]{khalil} shows that under suitable regularity of $\fav$, a Lyapunov function $V$ satisfying~\eqref{eq:V_av} exists if the average system is globally exponentially stable for each frozen $u_s$. Furthermore, such a $V$ exists globally if $\fav$ is globally Lipschitz (which is restrictive); otherwise, it exists on a compact set.
\begin{assumption}[Slow input]\label{ass_slow}
Assume that $u_s$ has finitely many discontinuities
on any bounded interval, is c\`adl\`ag, is $C^1$ between discontinuities, and the Euclidean norm $|\dot{u}_s(\cdot)|$ is Riemann
integrable between discontinuities.
\end{assumption}
The total variation (continuous $+$ discrete) of $u_s$ is defined.
\begin{definition}[Total variation~\citep{gao2018}]\label{def_tv}
Consider a signal $u_s$ on an interval $[t_1,t_2]$ with its discontinuities $d_1<d_2<\ldots< d_m$ such that $t_1 < d_1$ and $d_m \leq t_2$. The \emph{total variation} of $u_s$ on $[t_1,t_2]$ is defined as
\begin{equation}
\int_{t_1}^{t_2}\|du_s\| := \sum_{i=0}^m \int_{d_i}^{d_{i+1}}|\dot{u}_s(t)|dt + \sum_{i=1}^m|u_s(d_i^+) - u_s(d_i^-)|,
\end{equation}
where we set $d_0 := t_1$ and $d_{m+1} := t_2$.
\end{definition}
Note that at jump times $u_s(d^+) = u_s(d) \neq u_s(d^-)$ as $u_s$ is c\`adl\`ag---see Assumption~\ref{ass_slow}. Our main result is stated next.
\begin{theorem}[Stability of system~\eqref{eq:sys}]\label{theo_stab}
Consider system~\eqref{eq:sys} under Assumptions~\ref{ass_f},~\ref{ass_fast_1},~\ref{ass_fast_2},~\ref{ass_average}, and~\ref{ass_slow}. Assume that there exist $0 \leq \mu < \frac{c_1 c_3}{c_2 c_4}$ (where constants $c_i$ come from Assumption~\ref{ass_average}) and $\alpha \geq 0$ such that for every interval $[t_1,t_2]$, we have
\begin{equation}\label{eq:cond_tv}
\int_{t_1}^{t_2}\|du_s\| \leq \mu(t_2-t_1)+\alpha.
\end{equation}
For every $R > 0$, there exist $\e^\star,\rho, \lambda>0$ such that, for all $0<\e<\e^\star$, the solution $x(t)$ to system~\eqref{eq:sys} with $|x(0)| \leq R$ satisfies
\begin{equation}\label{eq:xstable}
|x(t)|
\leq \rho e^{-\lambda t}|x(0)|, \qquad \forall t\geq 0.
\end{equation}
\end{theorem}

\begin{remark}
While we assume that the general averaging is valid for all $(x,u_s) \in \RR^n \times \RR^m$, this is not always possible. For instance, with the toy example 
\begin{equation}
\dot{x}(t) = e^{-x(t) u_f(t/\e)}x(t), 
\qquad u_f(s) = s,
\end{equation}
the averaging~\eqref{eq:ave} can only be done with $x \geq 0$. In such cases, the basin of attraction characterized by $R$ in Theorem~\ref{theo_stab} cannot be taken arbitrarily large, but must be inside the region where the averaging is valid. We can accordingly relax Assumption~\ref{ass_average} from global stability into one with a basin that fits in the validity region of averaging. In any case, we can achieve stability that is semi-global with respect to the validity region of the general averaging and of stability of the frozen average system. This issue does not appear in earlier works such as~\cite{LibShimTAC25} where $u_f$ is periodic making the averaging independent of $(x,u_s)$. Such complicated cases will be treated in our future work. 
\end{remark}

\section{Proof of Theorem~\ref{theo_stab}}
This proof has two steps: (1) Define a coordinate change and derive the target dynamics, and (2) Prove stability in the new coordinates and come back to the original ones. Due to space constraints, we shorten some steps that can be found in~\cite{LibShimTAC25}---see Remark~\ref{rem_proof}.

\subsection{Change of Coordinates} 

Define 
\begin{equation}\label{eq:w}
w(x,u_s,\tau,\e) := \int_0^\tau h(x,u_s,u_f(s))e^{-\e(\tau-s)}ds.
\end{equation}
This function satisfies $w(0,\cdot,\cdot,\cdot) = 0$. Under Assumptions~\ref{ass_fast_1} and~\ref{ass_fast_2}, we deduce that there exists a class-$\K$ function $\gamma$ such that for any compact set $D \subset \RR^n \times \RR^m$, for all $(x,u_s,\tau,\e) \in D \times [0,+\infty) \times [0,+\infty)$, we have
\begin{align}
\e |w(x,u_s,\tau,\e)|&\leq \gamma(\e),\label{eq:khalil1}\\
\e \left\|\frac{\partial w}{\partial x}(x,u_s,\tau,\e)\right\| &\leq \gamma(\e),\label{eq:khalil2}\\
\e \left\|\frac{\partial w}{\partial u_s}(x,u_s,\tau,\e)\right\| &\leq \gamma(\e).\label{eq:khalil3}
\end{align}
These bounds are obtained as a generalization of~\cite[Page 416]{khalil}. Note that in~\cite{LibShimTAC25}, the boundedness of these functions in time is guaranteed by the periodicity of $u_f$, while here it is established by uniform bounds only in $\e$.
Consider a $C^1$ time-varying change of coordinates
\begin{equation}\label{eq:coc}
y := \Phi_{t,\e}(x) = x - \e w(x,u_s(t),t/\e,\e).
\end{equation}
This transformation is origin-preserving, i.e., $\Phi_{t,\e}(0) = 0$ for all $t,\e$. Reasoning in the same way as~\cite{ChoiShimSeo}, we deduce that there exists $\e^\star_1 > 0$ such that for each $(u_s,t,\e) \in \Gamma \times [0,+\infty) \times (0,\e^\star_1]$, this transformation is $C^1$ and bijective, and we denote its inverse by $x = \Phi_{t,\e}^{-1}(y)$. Moreover, we can show as in~\cite{LibShimTAC25} that $\underline{\Lambda}|y| \leq |x| \leq \overline{\Lambda}|y|$ with $\underline{\Lambda},\overline{\Lambda} \to 1$ as $\e^\star_1 \to 0$. Then, we can show that away from the discontinuities of $u_s$,
\begin{align*}
&\dot{y}(t) = \left(I - \e \frac{\partial w}{\partial x}(x(t),u_s(t),t/\e,\e)\right)\dot{x}(t) \\&{}- \e \frac{\partial w}{\partial u_s}(x(t),u_s(t),t/\e,\e)\dot{u}_s(t) - \e \frac{\partial w}{\partial \tau}(x(t),u_s(t),t/\e,\e)\frac{1}{\e}\\
& =  f(x(t),u_s(t),u_f(t/\e)) - \e \frac{\partial w}{\partial x}(x(t),u_s(t),t/\e,\e)\\&\quad{}\times f(x(t),u_s(t),u_f(t/\e))- \e \frac{\partial w}{\partial u_s}(x(t),u_s(t),t/\e,\e)\dot{u}_s(t)\\&\quad{} - h(x(t),u_s(t),u_f(t/\e)) + \e w(x(t),u_s(t),t/\e,\e)\\
& = \fav(y(t),u_s(t)) + \fav(x(t),u_s(t)) - \fav(y(t),u_s(t)) \\&\quad{}- \e \frac{\partial w}{\partial x}(x(t),u_s(t),t/\e,\e)f(x(t),u_s(t),u_f(t/\e)) \\&\quad{}- \e \frac{\partial w}{\partial u_s}(x(t),u_s(t),t/\e,\e)\dot{u}_s(t) + \e w(x(t),u_s(t),t/\e,\e).
\end{align*}
Similar to~\cite{LibShimTAC25}, using the Mean Value Theorem and differentiability of $\fav$ in $x$, we have
\begin{multline*}
\fav(x(t),u_s(t)) - \fav(y(t),u_s(t)) \\= \e F(\Phi_{t,\e}^{-1}(y(t)),y(t),u_s(t))w(\Phi_{t,\e}^{-1}(y(t)),u_s(t),t/\e,\e),
\end{multline*}
where the function $F$ is defined as
$$
F(x,y,u_s) := \int_0^1 \frac{\partial \fav}{\partial x}(\theta x + (1-\theta)y,u_s)d\theta.
$$
Hence, we obtain
\begin{subequations}
\begin{multline}\label{eq:ydot}
\dot{y}(t) = \fav(y(t),u_s(t)) + g_0(y(t),u_s(t),t,\e)\dot{u}_s(t)\\ + g_1(y(t),u_s(t),t,\e),
\end{multline}
where
\begin{align}
&g_0(y,u_s,t,\e) = - \e \frac{\partial w}{\partial u_s}(\Phi_{t,\e}^{-1}(y),u_s,t/\e,\e),\\
&g_1(y,u_s,t,\e)= \e F(\Phi_{t,\e}^{-1}(y),y,u_s)w(\Phi_{t,\e}^{-1}(y),u_s,t/\e,\e)  \nonumber\\&\qquad{}- \e \frac{\partial w}{\partial x}(\Phi_{t,\e}^{-1}(y),u_s,t/\e,\e)  f(\Phi_{t,\e}^{-1}(y),u_s,u_f(t/\e))\nonumber\\&\qquad{}+ \e w(\Phi_{t,\e}^{-1}(y),u_s,t/\e,\e).\label{eq:g1}
\end{align}
\end{subequations}
Similar to~\cite{LibShimTAC25}, we can show that for every compact $\Omega \subset \RR^n$, there exist $\delta_1,\delta_2 > 0$ such that, for every $t \geq 0$ where $\dot{u}_s(t)$ exists, every $\e \in (0,\e^\star_1]$, and every $y \in \Phi_{t,\e}(\Omega)$, we have
\begin{multline}\label{eq:bound}
|g_0(y,u_s(t),t,\e)\dot{u}_s(t) + g_1(y,u_s(t),t,\e)| \\\leq \delta_1 \gamma(\e) |y| + \delta_2 \gamma(\e) |y| |\dot{u}_s(t)|.
\end{multline} 
At a discontinuity $t$ of $u_s$, $y(t)$ experiences a jump (while $x(t)$ does not, or $x(t^+) = x(t^-)$) described by
\begin{align*}
y(t^+) &= \Phi_{t^+,\e}(x(t^+))= \Phi_{t^+,\e}(x(t^-)) \\& = \Phi_{t^+,\e} (\Phi_{t^-,\e}^{-1}(y(t^-))).
\end{align*}
Using the definition of $y$ and that $x(t^+) = x(t^-)$, we have
\begin{multline*}
y(t^+) - y(t^-)= -\e (w(x(t^-),u_s(t^+),t/\e,\e)\\ - w(x(t^-),u_s(t^-),t/\e,\e)).
\end{multline*}
Define the path $\kappa(r) := ru_s(t^+) + (1-r) u_s(t^-)$ parameterized by $r \in [0,1]$, which lies in $\Gamma$ thanks to convexity. By the Fundamental Theorem of Calculus and chain rule
\begin{align*}
&w(x(t^-),u_s(t^+),t/\e,\e) - w(x(t^-),u_s(t^-),t/\e,\e)\\
&= \int_0^1 \frac{d}{dr} w(x(t^-),\kappa(r),t/\e,\e)dr\\
&= \int_0^1 \frac{\partial w}{\partial u_s}(x(t^-),\kappa(r),t/\e,\e)(u_s(t^+) - u_s(t^-))dr\\
&= \left(\int_0^1 
\frac{\partial w}{\partial u_s}(x(t^-),\kappa(r),t/\e,\e)dr\right)
(u_s(t^+) - u_s(t^-)).
\end{align*}
Therefore, using~\eqref{eq:khalil3}, we obtain
\begin{multline}\label{eq:ypm}
|y(t^+) - y(t^-)|\leq \e \sup_{r \in [0,1]}\left\|\frac{\partial w}{\partial u_s}(x(t^-),\kappa(r),t/\e,\e)\right\| \\ 
\times|u_s(t^+) - u_s(t^-)| \leq \gamma(\e) |u_s(t^+) - u_s(t^-)|.
\end{multline}

\subsection{Stability in the New and Original Coordinates}
We define
\begin{equation}
  V(t):=V(y(t),u_s(t))
\end{equation}
 and study the evolution (flows and jumps) of $V(t)$ along $y(t)$ and $u_s(t)$. Away from the discontinuities of $u_s$, the time derivative of the Lyapunov function $V$ from Assumption~\ref{ass_average} along the system in the $y$-coordinates is given by 
\begin{align*}
&\dot{V}(y(t),u_s(t)) = \frac{\partial V}{\partial y}(y(t),u_s(t))(\fav(y(t),u_s(t)) \\&\qquad{}+ g_0(y(t),u_s(t),t,\e)\dot{u}_s(t) \\&\qquad{} + g_1(y(t),u_s(t),t,\e)) + \frac{\partial V}{\partial u_s}(y(t),u_s(t))\dot{u}_s(t).
\end{align*}
We get from Assumption~\ref{ass_average} and~\eqref{eq:bound} in a similar way as~\cite{LibShimTAC25} that
\begin{align*}
&\dot{V}(y(t),u_s(t))\leq -c_3 |y(t)|^2 + c_5\delta_1\gamma(\e)|y(t)|^2\\
&\qquad {} + c_4 |y(t)|^2|\dot{u}_s(t)|+ c_5\delta_2 \gamma(\e)|y(t)|^2|\dot{u}_s(t)| \\&\leq \left(-\frac{c_3}{c_2} + \frac{c_5}{c_1} \delta_1 \gamma(\e) +\left(\frac{c_4}{c_1} + \frac{c_5}{c_1} \delta_2 \gamma(\e)\right)|\dot{u}_s(t)|\right) \\&\qquad{}\times V(y(t),u_s(t)).
\end{align*}
Hence, for any interval $[t_1,t_2)$ not containing any discontinuity of $u_s(t)$, we have
\begin{multline}\label{eq:Vflow}
V(t_2) \leq \exp\bigg(\left(-\frac{c_3}{c_2} + \frac{c_5}{c_1} \delta_1 \gamma(\e)\right)(t_2 - t_1)\\+\frac{c_4}{c_1} + \frac{c_5}{c_1} \delta_2 \gamma(\e)\int_{t_1}^{t_2}|\dot{u}_s(t)|dt\bigg)V(t_1).
\end{multline}
At a discontinuity $t$ of $u_s$, we have $y(t^+) = \Phi_{t^+,\e}(x(t^-))$ and $u_s(t^+) \neq u_s(t^-)$. We decompose
\begin{multline}\label{eq:Vsplit}
V(t^+) - V(t^-)=V(y(t^+),u_s(t^+)) - V(y(t^-),u_s(t^-)) 
\\= V(y(t^+),u_s(t^+)) - V(y(t^-),u_s(t^+))
 \\+ V(y(t^-),u_s(t^+)) - V(y(t^-),u_s(t^-)).
\end{multline}
For the first term of~\eqref{eq:Vsplit}, proceeding in the same way as~\cite{LibShimTAC25} with the Mean Value Theorem and some subsequent inequalities with~\eqref{eq:ypm}, we deduce that there exist $\ell_1,\ell_2 > 0$ not depending on $\e$ such that
\begin{multline*}
V(y(t^+),u_s(t^+)) - V(y(t^-),u_s(t^+)) 
\\\leq (\ell_2 (\gamma(\e))^2 + \ell_1 \gamma(\e))|u_s(t^+) - u_s(t^-)|V(y(t^-),u_s(t^-)).
\end{multline*}
For the second term of~\eqref{eq:Vsplit}, from the Mean Value Theorem and~\eqref{eq:V_av}, there exists $\overline{u}_s = ru_s(t^+) + (1-r)u_s(t^-)$ for some $r \in [0,1]$, which is in $\Gamma$ thanks to convexity, such that
\begin{align*}
&V(y(t^-),u_s(t^+)) - V(y(t^-),u_s(t^-))\\&\leq  \left|\frac{\partial V}{\partial u_s}(y(t^-), \overline{u}_s)\right||u_s(t^+) - u_s(t^-)|\\&\leq c_4|y(t^-)||u_s(t^+) - u_s(t^-)|\\&\leq \ell_0|u_s(t^+) - u_s(t^-)|V(y(t^-),u_s(t^-)),
\end{align*}
where $\ell_0 := c_4/c_1$. Thus, we get that
\begin{equation}\label{eq:Vjump}
V(t^+) \leq \exp\left(\ell(\e)|u_s(t^+) - u_s(t^-)|\right)V(t^-),
\end{equation}
where $\ell(\e) = \ell_2 (\gamma(\e))^2 + \ell_1 \gamma(\e) + \ell_0$. Iteratively combining~\eqref{eq:Vflow} and~\eqref{eq:Vjump}, we get that over any interval $[t_1,t_2]$,
\begin{align*}
V(t_2) &\leq \exp\bigg(\left(-\frac{c_3}{c_2} + \frac{c_5}{c_1} \delta_1 \gamma(\e)\right)(t_2 - t_1)\\&\quad{}+\left(\frac{c_4}{c_1} + \ell_2 (\gamma(\e))^2 + \frac{c_5}{c_1} \max\{\delta_2,1\} \gamma(\e)\right)\\&\quad{}
\times\int_{t_1}^{t_2}\|du_s\|\bigg)V(t_1)\\ & \leq \exp\bigg(\left(-\frac{c_3}{c_2} + \frac{c_5}{c_1} \delta_1 \gamma(\e)\right)(t_2 - t_1)\\
&\quad {}+\left(\frac{c_4}{c_1} + \ell_2 (\gamma(\e))^2 + \frac{c_5}{c_1} \max\{\delta_2,1\} \gamma(\e)\right)\\&\quad{}\times(\mu(t_2 - t_1) + \alpha)\bigg) V(t_1),
\end{align*}
so that, since $0 \leq \mu < \frac{c_1 c_3}{c_2 c_4}$, we obtain exponential stability in the $y$-coordinates if $\e$ is sufficiently small. This stability is recovered in the $x$-coordinates in a similar way as in~\cite{LibShimTAC25}. Since the averaging is assumed to be valid for $(x,u_s)$ on the whole $\RR^n \times \RR^m$ and the exponential stability with frozen $u_s$ holds globally in $x$, we can take $R$ arbitrarily large. Theorem~\ref{theo_stab} thus follows.

\begin{remark}\label{rem_proof}
This proof resembles that in~\cite{LibShimTAC25}. The main modifications are due to general averaging replacing the periodic one:
\begin{itemize}[leftmargin=*,nosep]
\item Instead of bounding the norm of $w$ of the near-identity transformation and its partial derivatives by a constant (thanks to periodicity), here we have a modified $w$ (see~\eqref{eq:w}) and we bound $\e|w|$ together with the derivatives by a class-$\K$ function of $\e$ (see~\eqref{eq:khalil1}-\eqref{eq:khalil3});
\item There is an extra last term in $\dot{y}(t)$ dynamics caused by the new definition of $w$, which is contained in $g_1(y,u_s,t,\e)$ (see~\eqref{eq:g1}) and is bounded similarly.
\end{itemize}
\end{remark}

\section{Illustration: Switched System}
We consider a two-dimensional nonlinear switched system with dynamics
\begin{equation}\label{eq:sys_switch}
\dot x(t) = \frac{1}{1+|x(t)|}A_{\sigma(t)}x(t),
\end{equation}
and with four modes, where $\sigma(t)\in\{1,2,3,4\}$ is a slow-fast quasi-periodic switching signal defined below. The mode matrices $A_i$ ($i = 1,2,3,4$) are constructed so that
\begin{subequations}\label{eq:Aav}
\begin{align}
\frac{1}{2}(A_1 + A_2) & = \begin{pmatrix}
-0.3 & -20/3 \\
0.6  & -0.3
\end{pmatrix} =: A_{av,1},\\
\frac{1}{2}(A_3 + A_4) &= \begin{pmatrix}
-0.3 & -0.6 \\
20/3 & -0.3
\end{pmatrix} =: A_{av,2}.
\end{align}
\end{subequations}
All modes share the same equilibrium at the origin. Moreover, we pick some of these $A_i$ matrices to be non-Hurwitz, so that system~\eqref{eq:sys_switch} when fixed at such modes is unstable.

The switching law is designed to be the composition of slow switching between the two groups $\{1,2\}$ and $\{3,4\}$ and fast switching within each group, both of which are non-periodic. More formally, let $\e_s>0$ and $\e_f>0$ be tuning parameters for the slow and fast time scales (typically $\e_s \gg \e_f$), respectively, and define the quasi-periodic signals
\begin{subequations}
\begin{align}
r_s(t)& = \sin(t/\e_s) + \sin(\sqrt{2}t/\e_s),\\
r_f(t) &= \sin(t/\e_f) + \sin(\sqrt{5}t/\e_f).
\end{align}
\end{subequations}
Then, the slow group selector and the fast within-group selector are chosen respectively as
\begin{equation}
q_s(t)=
\begin{cases}
1, & r_s(t)\geq 0,\\
2, & \text{otherwise},
\end{cases} \quad q_f(t)=
\begin{cases}
1, & r_f(t)\geq 0,\\
2, & \text{otherwise}.
\end{cases}
\end{equation}
Finally, the mode $\sigma(t)\in\{1,2,3,4\}$ is defined by
\begin{equation}
\sigma(t)=
\begin{cases}
1, & q_s(t)=1,\ q_f(t)=1,\\
2, & q_s(t)=1,\ q_f(t)=2,\\
3, & q_s(t)=2,\ q_f(t)=1,\\
4, & q_s(t)=2,\ q_f(t)=2.
\end{cases}
\end{equation}
For analysis purposes, it is relevant to introduce the following switched system with two modes
\begin{equation}\label{eq:sys_switch_ave}
\dot{x} = \frac{1}{1+|x|}A_{av,1}x,\qquad \dot{x} = \frac{1}{1+|x|}A_{av,2}x.
\end{equation}
This new system~\eqref{eq:sys_switch_ave} is globally exponentially stable when staying in each mode, as can be shown with the quadratic Lyapunov function $V_i(x) = x^\top P_i x$ where each $P_i = P_i^\top > 0$ is solution to 
$$A_{av,i}^\top P_i + P_i A_{av,i} = -I, \qquad i = 1,2,
$$ 
which exists because each $A_{av,i}$ is Hurwitz. However, there is no common solution $P = P^\top > 0$ for both $i = 1,2$.

With our construction of the switched system~\eqref{eq:sys_switch}, when $\e_f$ is sufficiently small, the fast switching within each group yields average dynamics that are close to~\eqref{eq:sys_switch_ave}, i.e., the dynamics matrix is close to $A_{av,1}$ on intervals where $q_s(t)=1$, and close to $A_{av,2}$ on intervals where $q_s(t)=2$, while the slow switching between these two average dynamics is tuned by $\e_s$. The average dynamics will be exactly~\eqref{eq:sys_switch_ave} if the fast switching is periodic and the time spent between the two in-group modes is equal, due to the relation~\eqref{eq:Aav}. Moreover, there is a mode-dependent quadratic Lyapunov function for the average system~\eqref{eq:sys_switch_ave}, which guarantees global exponential stability when this system remains at each mode. This makes the average system~\eqref{eq:sys_switch_ave} satisfy Assumption~\ref{ass_average}. Next, we will simulate system~\eqref{eq:sys_switch} with different choices of $\e_s$, $\e_f$, and plot the trajectories in the state space with color indicating time.

\begin{figure}[ht]
\centering
\includegraphics[width=0.494\columnwidth,height=0.37\columnwidth]{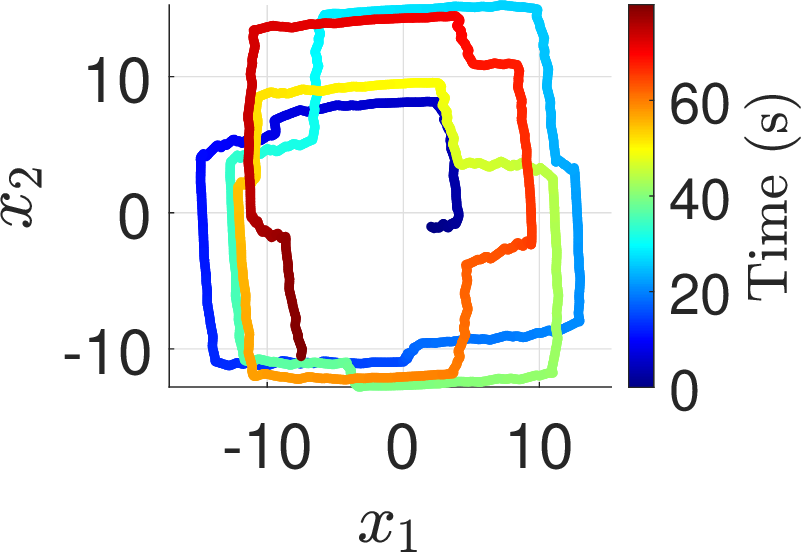}
\includegraphics[width=0.494\columnwidth,height=0.37\columnwidth]{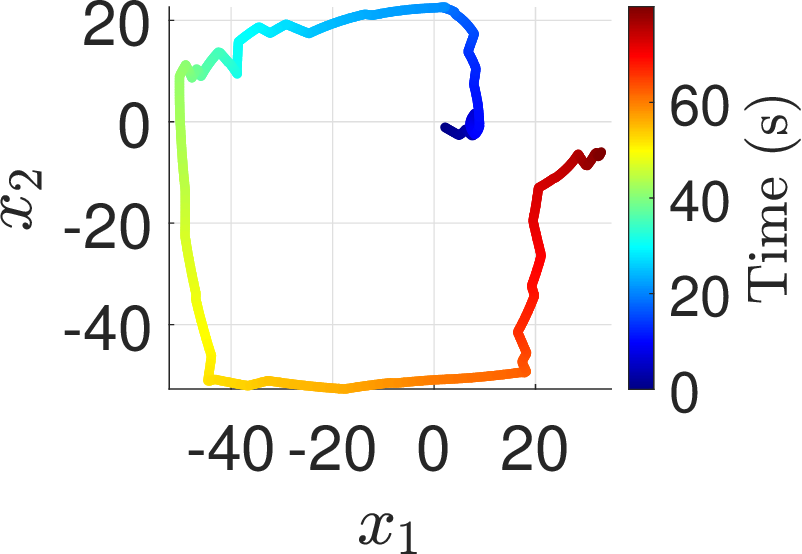}
\caption{Unstable state trajectories of system~\eqref{eq:sys_switch} with $\e_s = 1$, $\e_f = 0.1$ (left) and $\e_s = 4$, $\e_f = 1$ (right).\label{fig1}}
\end{figure}

\begin{figure}[ht]
\centering
\includegraphics[width=\columnwidth,height=0.84\columnwidth]{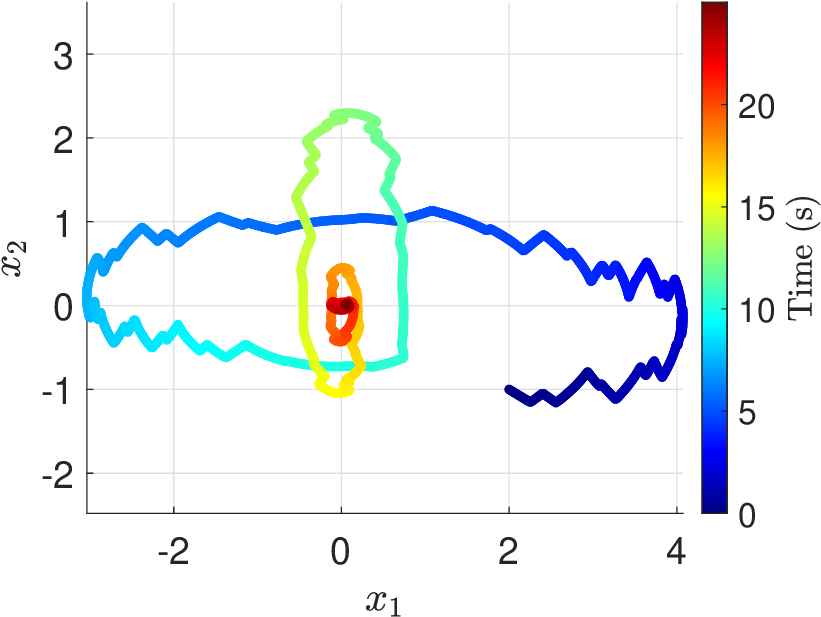}
\caption{Stable state trajectories of system~\eqref{eq:sys_switch} with $\e_s = 4$, $\e_f = 0.1$. \label{fig2}}
\end{figure}

Simulation results with $\e_s = 1$, $\e_f = 0.1$ and with $\e_s = 4$, $\e_f = 1$ shown in Fig.~\ref{fig1} indicate that the switched system~\eqref{eq:sys_switch} is unstable. In the first case, the slow switching is not slow enough, and so the stability of the average system~\eqref{eq:sys_switch_ave} when remaining at $A_{av,1}$ or $A_{av,2}$ does not carry over to system~\eqref{eq:sys_switch}. In the second case, the fast switching is not fast enough, and so the averaging cannot make system~\eqref{eq:sys_switch} behave similarly to the average dynamics~\eqref{eq:sys_switch_ave}. We then experiment with $\e_s = 4$ and $\e_f = 0.1$, which are respectively slow and fast enough, and stability is achieved as shown in Fig.~\ref{fig2}. This illustrates the result in our Theorem~\ref{theo_stab}, which is consistent with those in the literature of switched systems like~\cite{hespanha1999,shorten2007stability,gao2018} in terms of the \emph{slow} switching. Compared to those works, as stated in Section~\ref{sec:intro}, our handling of non-periodic fast variations allows us to incorporate both fast non-periodic switching and quasi-periodic variations as illustrated by this example. In this case, due to averaging, it is not important what the $A_i$ matrices are individually (or whether they are Hurwitz), as long as they satisfy~\eqref{eq:Aav} and the fast switching is fast enough.

\section{Conclusion}
This work presents sufficient conditions for the semi-global exponential stability of nonlinear dynamics with slow and fast time variations. Unlike related works, we consider the case where the fast variation is non-periodic, necessitating general averaging. Our result is illustrated using a switched system with slow and fast non-periodic switching.

Future work will consider the more general case of \emph{moving} equilibria as well as related applications, such as HIV treatment~\citep{ChoiShimSeo} or PWM-based control~\citep{pwm2025}. Conditions related to general averaging, such as for $\fav$ in~\eqref{eq:ave} to be $C^1$, which we assume to hold in this paper, will also be studied.

\begin{ack}
The work of G. Q. B. Tran and D. Liberzon is supported in part by the AFOSR MURI FA9550-23-1-0337 and NSF CMMI-2452534 grants. The work of H. Shim is supported in part by the National Research Foundation of Korea (NRF) grant funded by the Korea government (MSIT) (No. RS-2022-00165417).
\end{ack}

\bibliography{ref} 
\end{document}